\documentclass[journal=jacsat,manuscript=article]{achemso}

\usepackage[version=3]{mhchem} 
\usepackage{color}
\usepackage{graphicx}
\usepackage{dcolumn}
\usepackage{bm}
\usepackage{multirow}

\newcommand{\fref}[1]{Figure~\ref{#1}} 
\newcommand{\tref}[1]{Table~\ref{#1}} 
\newcommand{\eref}[1]{Eqn.~\ref{#1}} 

\newcommand{\pka}{p$K_{\rm a}$\,}
\newcommand{\dpka}{$\Delta$p$K_{\rm a}$\,}

\author{Shivani Verma}
\affiliation{Department of Chemistry, Indian Institute of Technology Kanpur, Kanpur - 208016, India}

\author{Nisanth N. Nair}
\affiliation{Department of Chemistry, Indian Institute of Technology Kanpur, Kanpur - 208016, India}
\email{nnair@iitk.ac.in}

\title
  {Computational Study of p\boldmath$K_{\rm a}$\ shift of Aspartate residue in Thioredoxin: Role of Configurational Sampling and Solvent Model}

\abbreviations{FEP, TI, TI-dAFED, GB}

\begin{document}

\begin{tocentry}
\begin{figure}[H]
\centering
    \includegraphics[width=1.0\linewidth]{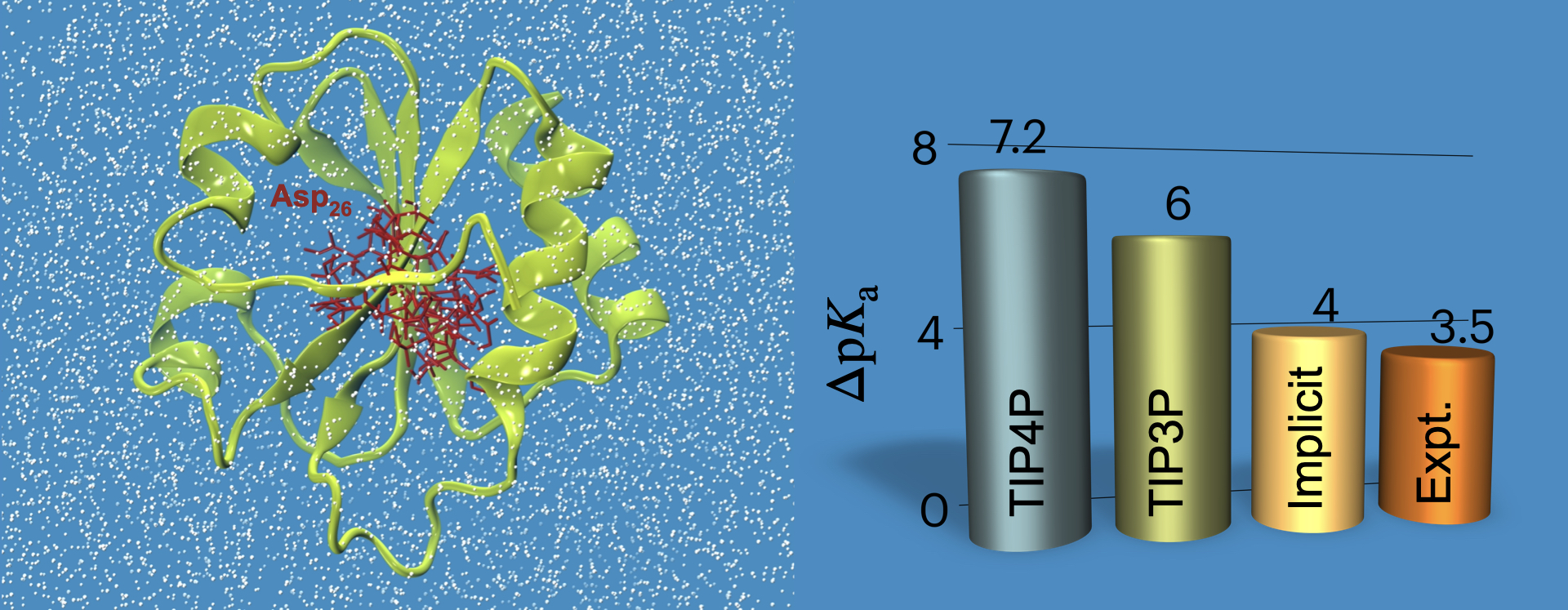}
\end{figure}
\end{tocentry}

\begin{abstract}
Alchemical free energy calculations are widely used in predicting \pka, and binding free energy calculations in biomolecular systems. 
These calculations are carried out using either Free Energy Perturbation (FEP) or Thermodynamic Integration (TI).
  %
Numerous efforts have been made to improve the accuracy and efficiency of such calculations, especially by boosting conformational sampling.
In this paper,  we use a technique that enhances the conformational sampling by temperature acceleration of collective variables for alchemical transformations and applies it to the prediction of \pka of the buried Asp$_{26}$ residue in thioredoxin protein.
  %
  %
We discuss the importance of enhanced sampling in the \pka calculations.
  %
The effect of the solvent models in the computed \pka values is also presented.
\end{abstract}

\section{Introduction}
Molecular dynamics (MD) is widely employed in calculating free energy differences between different molecular conformational states and free energy changes along physio-chemical processes in the condensed phase.\cite{tuckerman,Straatsma1992,Hansen2014,Beveridge1989,Frenkel}
Free energy calculations based on Thermodynamic Integration (TI)\cite{ti} and Free Energy Perturbation (FEP)\cite{fep} have been applied to a wide spectrum of problems in chemistry, and biology\cite{Kollman:ChemRev:93,Chipot:07,Gunsteren2010,Hage2018}, like drug discovery,\cite{Friesner2017,Sherman:JCIM:2017,Mobley:JCP:2012,Chodera:COSB:2011,Song2020} ligand binding in proteins\cite{Simonson:ACR:2002,Chipot:JCTC:2013,Gilson:ARB:2017,Panday2019}, identifying protonation states of ionizable residues through pK$_{\rm a}$
calculations,\cite{Simonson:JACS:2004,Jorgensen:JCTC:2014} conformational free energy differences,\cite{Straatsma:JCP:1989,Cuendet2018,He2018} and computing solvation free energies.\cite{Mobley:JCAMD:2010,Procacci2019,Deng2015}.
%
In these methods, free energy differences are calculated by introducing some non-physical intermediate states between two physically relevant states. 
When applied to condensed matter systems, the predictive power of these methods is affected by the slow convergence in the free energy estimates, mainly due to the drastic environmental changes while moving from one state to the other.
Systems get trapped in high-energy metastable states during the simulation resulting in poor conformational sampling.

This issue is addressed by combining the alchemical methods with enhanced sampling MD techniques.
Along these lines, FEP/TI combined with umbrella sampling,\cite{FEP-US,Ngo2021,Leitgeb2005} 
 TI-driven Adiabatic Free Energy Dynamics (dAFED),\cite{dafed,tidafed} FEP combined with Hamiltonian Replica Exchange Molecular Dynamics,\cite{Adrian2011}  FEP combined with solute tempering replica exchange and other global tempering methods\cite{Khavrutskii2010,Wang2015,Roux2018,Wang2019}, simulated scaling method for localized enhanced sampling,\cite{Yang2007} and thermodynamic integration with enhanced sampling (TIES)\cite{Coveney2017}
were proposed by various authors.
%

%
Amongst them, the TI-driven Adiabatic Free Energy Dynamics method is particularly interesting.
In this method, TI is done along with an enhanced sampling of collective variables (CVs) in the framework of dAFED in which a set of adiabatically decoupled auxiliary variables are coupled with the CVs.
A high temperature of the auxiliary variables is used to enhance the sampling of the CV space.
Auxiliary variables are harmonically coupled to the CVs, and for maintaining adiabatic decoupling, auxiliary variables are
assigned high masses.
%
The dAFED-based sampling can be further enhanced by biasing all or a subset of collective variables.\cite{Tuckerman:UFED,tass}
%

%
An alternative approach for TI/FEP is the $\lambda$-dynamics method, where the perturbation parameter $\lambda$ is treated as a dynamic variable\cite{lambda-dynamics}.
The original version has applied umbrella sampling\cite{umbrella} on the order parameter $\lambda$.
This method is further improved by combining it with enhanced sampling methods like metadynamics\cite{metadynamics}, named as $\lambda-$metadynamics\cite{lambda-metadynamics}.
The original $\lambda$-dynamics methodology was implemented for
modeling multiple substituents at a single site on a common
ligand framework. 
This technique has been combined with other CV-based biasing techniques, such as Local Elevation Umbrella Sampling.\cite{Bieler2014,Bieler2015,Hahn2020}
The improved version of this method, named multi-site $\lambda$-dynamics,\cite{MS-lambda-dyanmics,Hayes2017,Brooks2022} enables multiple substituents at multiple sites on a common ligand core.
$\lambda$-dynamics approach has various applications in studying relative protein stability and ligand binding\cite{lambda-dyanmics:review}.
In recent years, with the advances in machine learning approaches,  active learning protocols have been combined with alchemical methods to screen novel drug candidates.\cite{Groot2022}
Single-step FEP techniques like Enveloping Distribution Sampling and variants are also 
gaining attention.\cite{Han1992,Perthold2020,Knig2021}

The protonation state of ionizable amino acid residues is dictated by their interactions with the rest of the protein environment and the surrounding solvent.\cite{Isom2010}
The protonation state of the side chains can influence the structure of the proteins and their functions.\cite{Aghera2012}
\pka measurements provide valuable information about the protonation states of residues within the protein. 
Ionizable amino acids buried in the interior of proteins can have a substantial shift in its \pka relative to that in solution.
%
%
%
Determining the protonation states of the active site residues is critical for predicting the mechanism of enzymatic reactions.
\textit{Escherichia coli} thioredoxin, a soluble protein with 108 amino acids, is involved in various redox and regulatory activities.\cite{Holmgren1975}
In the active site of the thioredoxin, Asp$_{26}$ is buried in the hydrophobic core close to the redox-active disulfide residue and is known to play a critical role in the function of thioredoxin. 
Several computational studies have already reported the values of \dpka of Asp$_{26}$ of the protein and experimental measurement of \dpka is available.\cite{Sun2017,Adrian2011,Ji2008,Simonson:JACS:2004,Langsetmo1991,Dyson1991}
A large shift in \pka is reported for this system.\cite{Langsetmo1991,Dyson1991}
Thus this has been considered to be an ideal system for testing alchemical methods for \pka calculations. 
%
%
Simonson {\em et.~al.}~\cite{Simonson:JACS:2004} 
%
%
reported a $\Delta \Delta F$, which is the relative protonation free energy of Asp$_{26}$ residue in protein compared to the isolated Asp residue in water, to be {9.1 $\pm$ 4.1}~kcal~mol$^{-1}$.
Later, Meng~{\em et.~al.}~\cite{Adrian2011} used Hamiltonian Replica Exchange Molecular Dynamics combined with free energy perturbation.
%
%
The authors find that the replica exchange simulations boosted the conformational sampling, and the computed free energy change is in excellent agreement with the experimental data.
%
%
Ji~{\em et.~al.}~\cite{Ji2008} used polarized protein-specific charges (PPCs) to successfully reproduce the experimental \pka of thioredoxin in {explicit solvent} TI calculations.
%
%
%
Martinez et. al.\cite{Gomez2019} have shown that considering different protein conformations and polarization is critical for predicting the experimental \pka shift. 
%
%

In this paper, the TI-dAFED method is used to compute the \pka shift of Asp$_{26}$ in Escherichia coli thioredoxin.
We aim to probe the effect of boosting the conformational sampling in the \pka shift of Asp$_{26}$, mainly
considering that the residue is located within a hydrophobic core of the thioredoxin protein.
Further, solvent molecules can directly interact with the Asp$_{26}$ residue, making the \pka calculations challenging.
%
Explicit and implicit solvent simulations were performed to validate the results in the different solvent environments.
%
%
\section{Theory and Method}

\subsection{Thermodynamic Integration (TI)}
In the TI method,  potential energy is defined as,
\begin{eqnarray}
\label{eqn:U:TI}
    U(\mathbf R, \lambda)= f(\lambda) U_{\rm A}(\mathbf R)  + 
g(\lambda) U_{\rm B}(\mathbf R) 
\end{eqnarray} 
where $U_{\rm A}$ and $U_{\rm B}$ are the potential energy functions of the states A and B, respectively,
 ${\bf R}$ is the set of all atomic coordinates, and $\lambda$ is a parameter such that $\lambda \in [0,1]$.
Here, $f(\lambda)$ and $g(\lambda)$ are some functions of $\lambda$ such that $\lambda=0$ corresponds to state A, i.e., $U\equiv U_{\rm A}$, and $\lambda=1$ corresponds to state B. 
Any value of $\lambda$ between 0 and 1 corresponds to an intermediate state.
%
%
%
%
%
The free energy derivative with respect to $\lambda$ has the form
\begin{equation}
\label{eqn:dfdl}
    \left (\frac{\partial F}{\partial \lambda} \right )_{N,V,T} = \left \langle \frac{\partial U}{\partial \lambda}  \right \rangle \, 
\end{equation}
which can then be integrated to compute $\Delta F$:
\begin{equation}
 \Delta F = F_{\rm B} - F_{\rm A} = \int_{0}^1 d \lambda \left \langle \frac{\partial U}{\partial \lambda}  \right \rangle_\lambda \enspace . 
\end{equation}
In the above, the brackets $\left <  ... \right >$ represent  ensemble average in the canonical ensemble using the
potential $U(\lambda)$.
When $f(\lambda)=(1-\lambda)$  and $g(\lambda)=\lambda$, then,
\begin{equation}
\label{eqn:dudl_linear}
\left \langle \frac{\partial U}{\partial \lambda}  \right \rangle_\lambda =  \left \langle U_B - U_A \right \rangle_\lambda \enspace .
\end{equation}
In our calculations, only the electrostatic potential is changed while going
from Asp$_{26}$-H to Asp$_{26}^-$ with the change of $\lambda$ from 0 to 1.

\subsection{Thermodynamic Integration Driven-Adiabatic Free Energy Dynamics (TI-dAFED)}
In Temperature Accelerated Molecular Dynamics (TAMD)\cite{tamd} 
and in d-AFED\cite{dafed}, an extended Lagrangian is used:\cite{tuckerman,Awasthi2018}
\begin{eqnarray}
\label{eqn:TAMD:lag}
\mathcal L_{\rm TAMD/d-AFED}(\mathbf R, \dot{\mathbf R}, \mathbf z,\dot{\mathbf z}) = \mathcal L_{0}(\mathbf R, \dot{\mathbf R}) + \sum_{\alpha=1}^{n} \frac {1}{2} \mu_{\alpha} \dot{\mathbf z}_{\alpha}^2 - \nonumber \sum_{\alpha=1}^{n} \frac{k_{\alpha}}{2} ( q_{\alpha}(\mathbf R) - z_{\alpha})^2  \enspace
\end{eqnarray}
where $\mathcal L_{0}(\mathbf R, \dot{\mathbf R})$ is the original Lagrangian of the system,
$n$ is the number of CVs, $\mu_\alpha$ is the mass of the auxiliary degrees of variables $\left \{ z_\alpha \right \}$, and $k_\alpha$ is the
 coupling constant which determines the strength of the coupling between $\left \{ z_\alpha \right \}$ and 
 the CVs $\left \{ q_\alpha \right \}$.
%
The temperature of the auxiliary variables is kept much higher than the physical degrees of freedom.
This is achieved by coupling two different thermostats to these degrees of freedom.
%
The masses, $\left \{ \mu_\alpha \right \}$, are taken much higher than the atomic masses to maintain an 
adiabatic decoupling between the auxiliary and the physical degrees of freedom.
The high temperature of the auxiliary variables boosts the sampling of the CVs, which in turn helps the system
to explore the phase space efficiently.

In TI-dAFED simulations,\cite{tidafed} the Lagrangian $\mathcal L_0$ is composed of the potential energy $U(\mathbf R,\lambda)$ as given in \eref{eqn:U:TI}.
This allows us to enhance the exploration of the CV space while performing the TI simulations.
%
Appropriate reweighting factors are required to recover
the free energy differences, as shown below:
\begin{equation}
\label{eqn:deltaF:AFED}
    \Delta F = \int_{0}^{1} d\lambda \,   \int d \mathbf z  \, \, \left <  \frac{\partial U}{\partial \lambda}  \right >(\mathbf z;\lambda) \, \, A_{\lambda}(\mathbf z) \enspace ,
\end{equation}
where 
\begin{eqnarray} 
A_{\lambda}(\mathbf z)=  \frac{ \exp \left [-\beta {\phi}(\mathbf z)  \right ] }
{\int d\mathbf z  \exp \left [ -\beta {\phi}(\mathbf z)  \right ]  }
\enspace ,
\end{eqnarray}
and
\begin{equation}
    \phi(\mathbf z) = - \frac{1}{\beta_z} \ln P(\mathbf z)   \enspace .
\end{equation}
Here, $P(\mathbf z)$ is the probability distribution of auxiliary variables at temperature $T_z$. 
The temperature of the auxiliary variables $T_z$ is much higher than the physical temperature $T$, and 
$\beta_z = (k_{\rm B}T_z)^{-1}$ and $\beta = (k_{\rm B}T)^{-1}$, where $k_{\rm B}$ is the Boltzmann constant. 
The reweighting factors $A_\lambda(\mathbf z)$ are computed by a post-processing script on the 
bins created within the CV space.
In \eref{eqn:deltaF:AFED}, we require $\left <  \frac{\partial U}{\partial \lambda}  \right >(\mathbf z)$ on the same CV-bins, which in turn is computed by binning the $ \left (\frac{\partial U}{\partial \lambda}  \right ) (t)$ 
from the simulations, followed by local averaging on every bin.
%
\subsection{pK$_{\rm a}$ Shift Calculations}
\begin{figure}[htpb]
\centering
\includegraphics[width=0.8\textwidth]{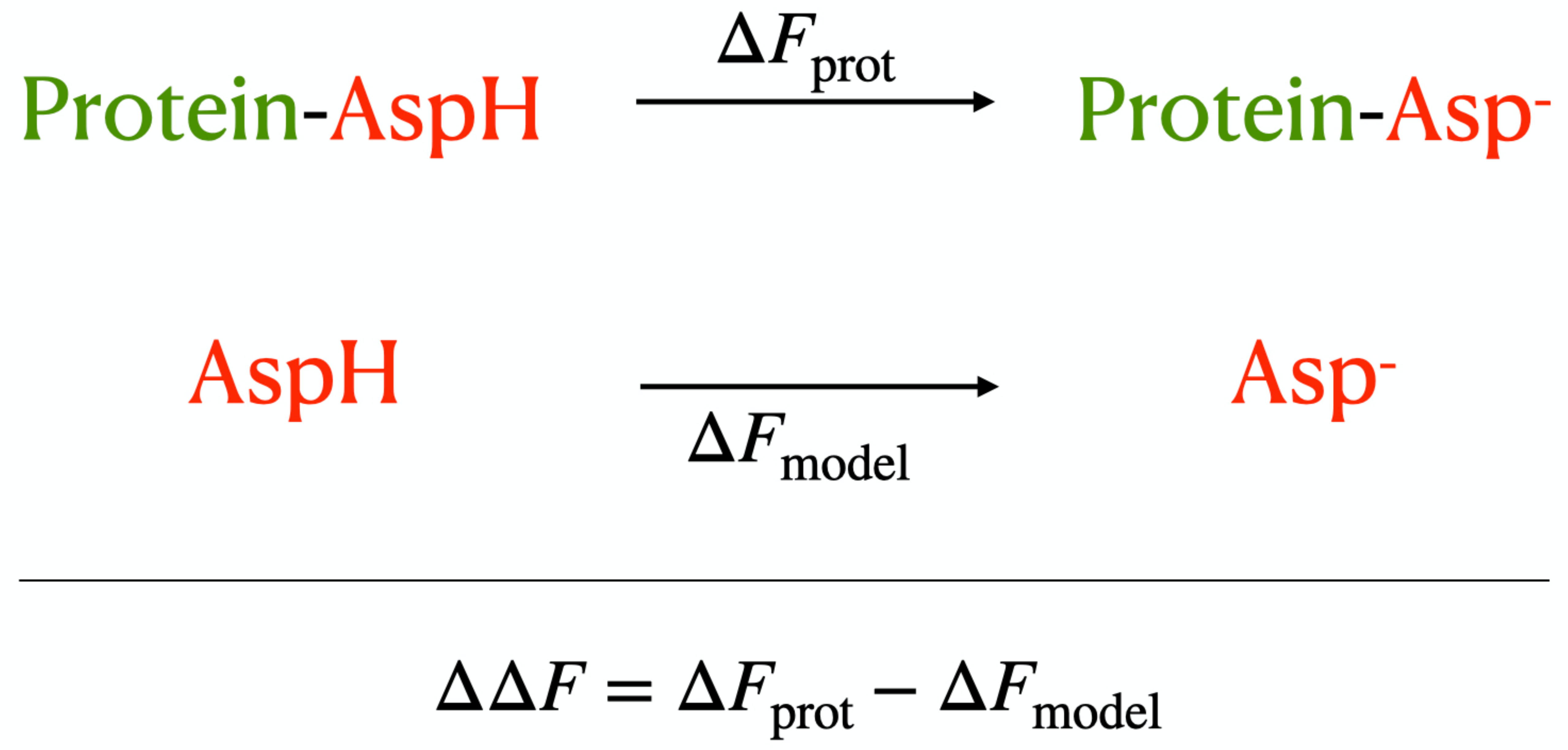}
    \caption{\label{fig:thermodynamic_cycle} Alchemical transformations involved in the calculation of \pka shifts ($\Delta$\pka) of Asp in thioredoxin are given. The first transformation shows the deprotonation of Asp residue in the protein environment, while the second transformation shows the deprotonation of Asp in water. $\Delta F$ values computed for these two alchemical transformations are used to compute the shift in $\Delta F$, i.e., $\Delta \Delta F$, which in turn is used to compute $\Delta$\pka.}
\end{figure}
\pka shift of aspartic acid (Asp) residue in the oxidized form of thioredoxin  is computed 
using $\Delta \Delta F$, which is the difference in free energy change for converting Protein-AspH to Protein-Asp$-$ ($\Delta F_{\rm prot}$)
and  free energy change for converting AspH to Asp$-$ ($\Delta F_{\rm model}$) in solution 
(\fref{fig:thermodynamic_cycle}):
\begin{eqnarray}
\label{pka}
    \Delta \mathrm{p}K_{\rm a} &=& \mathrm{p}K_{\rm a}\,(\mathrm{prot}) - \mathrm{p}K_{\rm a}\,(\mathrm{model}) \nonumber  \\
    & = & \frac{1}{2.303\, k_B T}  \left [ \Delta F_{\mathrm{prot}}  - \Delta F_{\mathrm{model}} \right ] \nonumber \\
    &=& \frac{1}{2.303\, k_B T} \Delta \Delta  F
\end{eqnarray}
Conversion of protonated Asp to deprotonated Asp is an alchemical change, as the proton disappears 
during this transformation.
Such transformations are performed in solvated protein and ligand systems using TI and TI-dAFED methods.
%
%
%
%
%
%
%

\subsection{Computational Setup}

Asp model system was constructed using  2\textit{N}-acetyl-1\textit{N}-methyl-aspartic acid-1-amide.
The protein structure was constructed from the PDB ID:2TRX.\cite{2trx:PDB}
Protonation states of the residues except for Asp$_{26}$ of the protein were set for pH=7.5.
N$_\delta$ and N$_\epsilon$ of His$_{6}$ is taken in the protonated state.
%
%
All calculations are done in the CUDA-enabled AMBER-18 PMEMD software \cite{amber18,AMBER_GPU,AMBER_TI_GPU1,AMBER_TI_GPU2} patched with PLUMED 2.6.1.\cite{plumed2.2.3}
The AMBER ff99SB force-field \cite{ff99SB} is used for all the simulations.
The SHAKE algorithm\cite{SHAKE} is used to constrain the covalent bonds with H-atoms.
The Langevin thermostat, as available in AMBER-18, was used to control the temperature of the system at 300~K.

%
We have considered $\lambda=0$ as Asp$_{26}$-H (protonated) state and $\lambda=1$
 as Asp$_{26}^-$ (deprotonated) state.
Partial charges for protonated and deprotonated states are taken from the earlier work.~\cite{Simonson:JACS:2004}
As $g(\lambda)$ and $f(\lambda)$ are linear functions of $\lambda$, intermediate
states are obtained by linearly interpolating the potential energy function.
We took 12 $\lambda$ points from 0.0 to 1.0, with a gap of 0.1 and an extra point at 0.05.

We performed implicit and explicit solvent MD simulations. 
Explicit water simulations are performed with TIP3P\cite{TIP3P} and TIP4P\cite{TIP4P} water models.
The initial box size for the explicit solvent simulations 
was 55$\times$60$\times$62\AA$^3$ and 32$\times$34$\times$29\AA$^3$
while simulating the solvated protein and the solvated model systems, respectively.
%
The protein and the model systems contained 4783 (4697) and 670 (662) water molecules, respectively, while using the 
TIP3P (TIP4P) force field.
%
The Onufriev, Bashford, and Case generalized Born implicit solvent approach \cite{Onufriev2002} was used for the implicit solvent 
simulations.
No counter-charges were present while doing the implicit solvent calculations.

For the case of explicit solvent simulations, we ran 2~ns of $NPT$ ensemble simulations until the density of the system was equilibrated. 
We performed 20~ns of $NVT$ equilibration for both implicit and explicit solvent models and all the $\lambda$ windows.
%
Starting structure for all other $\lambda$ values was taken from the equilibrated structure of the preceding $\lambda$ simulation. 
%
%
The production runs were for 100~ns for all the $\lambda$ windows.
Particle Mesh Ewald method \cite{PME} is used for calculating long-range interactions in 
all the explicit solvent simulations.
The frictional coefficient for the Langevin thermostat was taken to be 1 $\mathrm{ps}^{-1}$, and a time-step of 1~fs was used.
%
%
%
In the case of implicit solvent, the frictional coefficient for Langevin dynamics was taken as 5~$\mathrm{ps}^{-1}$,
and  2~fs time-step was used.
Berendsen barostat was used for the $NPT$ simulations\cite{Berendsen:barostat}.
%
The trapezoidal method was used for the numerical integrations concerning TI calculations.

\begin{figure}[htpb]
\centering
\includegraphics[width=0.8\textwidth]{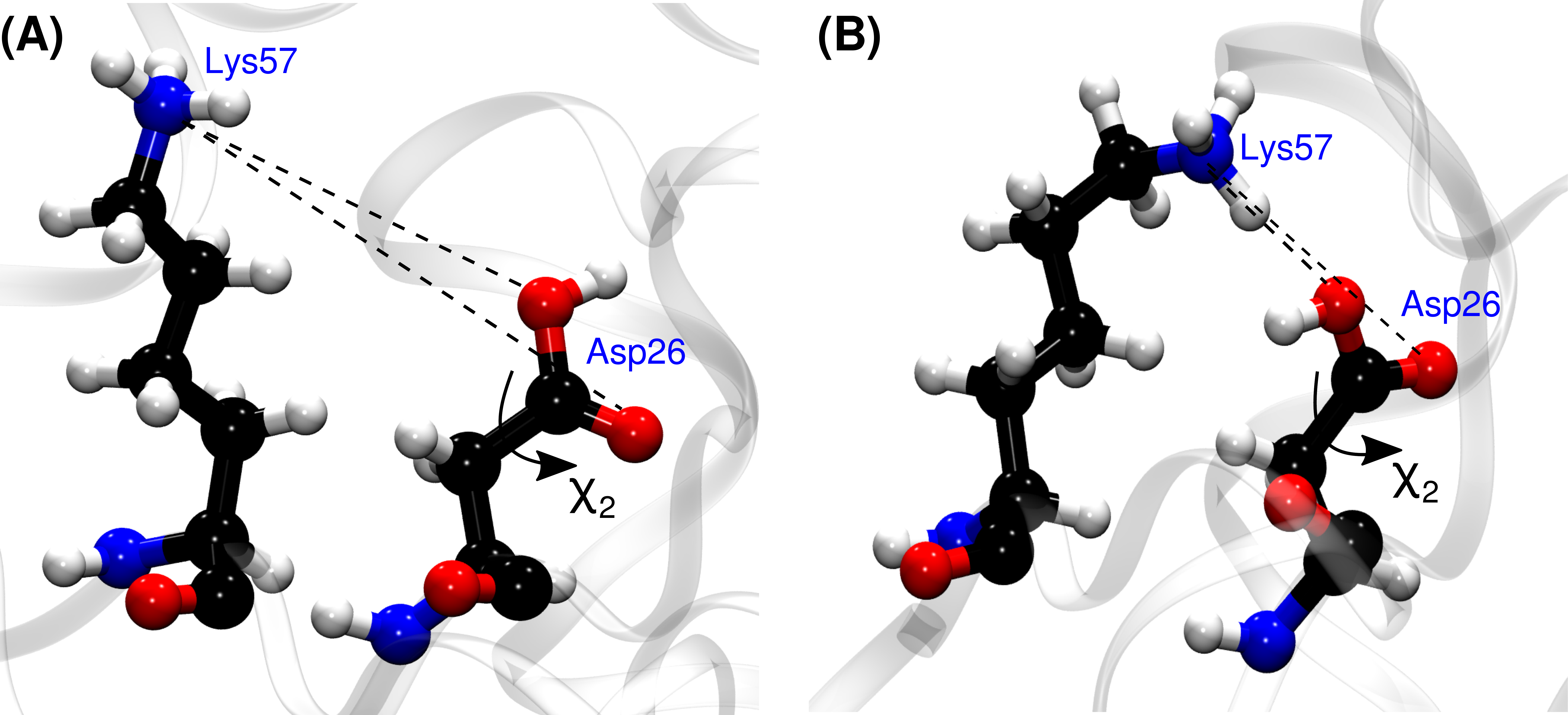}
    \caption{\label{fig:cv} 
     Snapshots showing two conformational states of Asp$_{26}$ in the protein. 
     The CV $\chi_2$ is labeled, and the two distances used as CVs are indicated by dotted lines. Atom colors: C (black), O (red), N (blue), H (white). 
   }
\end{figure}
%
%
Three collective variables were used to enhance various orientations of Asp$_{26}$ in protein at different values of $\lambda$ (see \fref{fig:cv}): (1) $\chi_2$ dihedral of Asp$_{26}$, (2) $d[\mathrm{Asp}_{26}\, \mathrm{O}_{\delta 1}-\mathrm{Lys}_{57} \, \mathrm{N}_\zeta]$, and (3) $d[\mathrm{Asp}_{26}\, \mathrm{O}_{\delta 2}-\mathrm{Lys}_{57} \, \mathrm{N}_\zeta]$.
While the first CV enhances the rotation about the dihedral $\chi_2$, the second, as well as the third CVs, boost formation and breakage of hydrogen-bonding interactions with Lys$_{57}$.
Real collective variables were coupled with extended CVs by a restraining potential with a spring constant of $1.2 \times 10^3$~kcal~mol$^{-1}$~rad$^{-2}$ for $\chi_2$ dihedral CV, $2.4 \times 10^3$ kcal~mol$^{-1}$nm$^{-2}$ for the other two CVs.
%
The masses for the three auxiliary variables were 
{50~a.m.u.~{\AA}$^2$~$\rm rad^{-2}$}, 
{266~a.m.u.}, and
{266~a.m.u.}, 
respectively.
The auxiliary variables coupled to the CVs were thermostatted to 1200~K using a Langevin thermostat.
It was found that the above parameters were sufficient to obtain a slow diffusion of the auxiliary variables $\{z_\alpha \}$ with respect to real-coordinates $\{q_\alpha \}$, and that $\{q_\alpha \}$ follows $\{z_\alpha \}$.
The average temperature of the auxiliary variables and the physical variables remained close to the 
target temperature.
%
%
\section{Results and Discussions}
\subsection{Implicit Solvent Simulation}

\begin{table}[H]
\caption{\label{t:deltaF} Computed $\Delta F_{\rm prot}$, $\Delta F_{\rm model}$, and $\Delta \Delta F$ using various methods and literature data are listed. 
The free energy values are in kcal/mol. 
All simulations were carried out for 100~ns per window.
%
}
\begin{tabular}{|c | c | c | c|}
\hline \hline 
Method & $\Delta F_{\rm prot}$ & $\Delta F_{\rm model}$ & $\Delta \Delta F = \Delta F_{\rm prot} - \Delta F_{\rm model}$ \\ \hline \hline
 TI/Implicit                  & -56.7 $\pm$ 0.9 & -62.0 $\pm$ 0.3 & 5.3 $\pm$ 0.9 \\ \hline
TI-dAFED/Implicit       & -56.7 $\pm$ 1.0 & -62.2 $\pm$ 0.5 & 5.5 $\pm$ 1.1 \\ \hline
 TI/TIP3P                   & -66.9 $\pm$ 2.8 & -75.1 $\pm$ 2.8 & 8.2 $\pm$ 4.0 \\ \hline
TI-dAFED/TIP3P       & -66.2 $\pm$ 3.1 & -74.5 $\pm$ 2.9 & 8.3 $\pm$ 4.2 \\ \hline
 TI/TIP4P                  & -70.9 $\pm$ 2.7 & -81.3 $\pm$ 2.9 & 10.4 $\pm$ 4.0 \\ \hline
TI-dAFED/TIP4P       & -70.7 $\pm$ 3.2 & -80.6 $\pm$ 3.0 & 9.9 $\pm$ 4.4 \\ \hline
Literature data: Ref.\cite{Simonson:JACS:2004}$^a$ & -66.0 $\pm$ 3.9 &  -75.1 $\pm$ 1.1 &  9.1 $\pm$ 4.1 \\ 
     \phantom{Literature data:} Ref.\cite{Adrian2011}$^b$ & -54.27 $\pm$ 0.22 &  -59.68 $\pm$ 0.08 &  5.41 $\pm$ 0.23 \\ \hline 
Experiment\cite{Langsetmo1991}                 &                  &              &       4.8 \\ \hline 
\end{tabular}\\
{$^a$ Conventional TI/Explicit: Ref.~\cite{Simonson:JACS:2004}} \\
{$^b$ FEP+H-REMD/Implicit: Ref.~\cite{Adrian2011}} \\
\end{table}

\begin{figure}[htpb]
\centering
\includegraphics[width=0.8\textwidth]{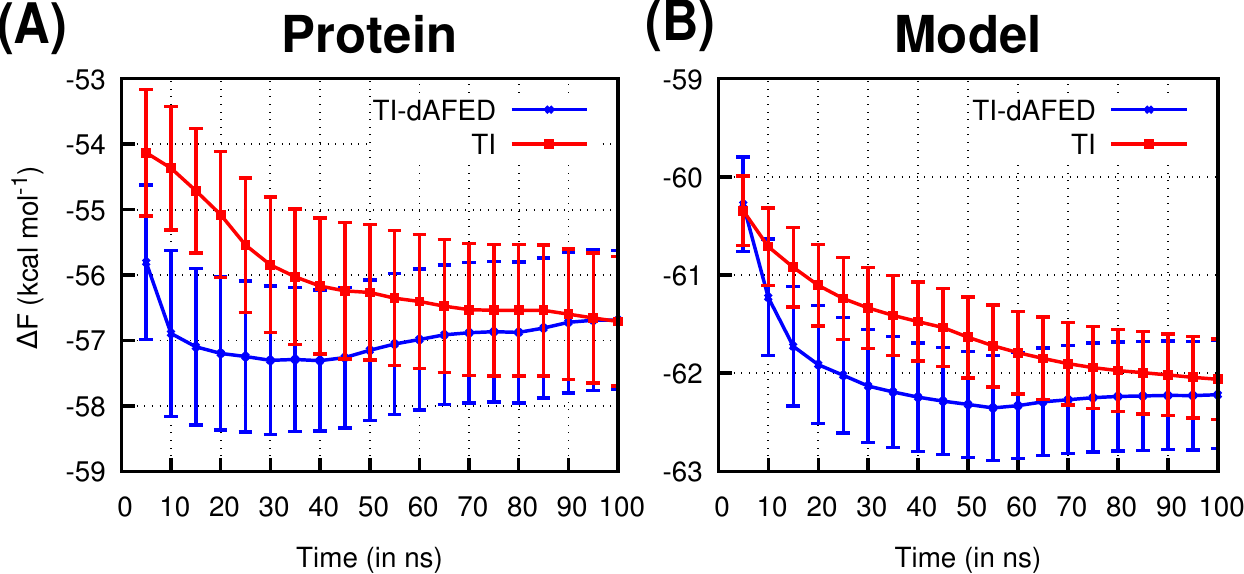}
    \caption{\label{fig:conv_implicit} 
    Convergence of $\Delta F$ in simulations using the implicit sol model for the (A) protein and the (B) model system. Results of TI-dAFED (blue) and TI (red) are presented. Error bars are also shown.
    }
\end{figure}

At first, we are presenting the data of TI calculations using the implicit solvent model.
The free energy differences $\Delta F$ were calculated for protein and model as discussed in Section~2 of the manuscript.
To check the convergence of the free energy estimate, we monitored $\Delta F$ as a function of simulation time (\fref{fig:conv_implicit}).
In the case of the model and the protein, $\Delta F$ has converged within 100~ns per $\lambda$ window.
%
%
Table~\ref{t:deltaF} has the converged values of $\Delta F_{\rm prot}$, $\Delta F_{\rm model}$, and $\Delta \Delta F$.
The same set of calculations was repeated using the TI-dAFED.
The results of both conventional TI and TI-dAFED are in excellent agreement with the experimental\cite{Langsetmo1991}
value and the previous simulation data using an implicit solvent.\cite{Adrian2011}
From \fref{fig:conv_implicit}, one may conclude that TI-dAFED has better convergence than TI; however, these differences were not substantial considering the error in the estimates.
For conventional TI calculations, $\Delta F_{\rm prot}$ converges at about 30~ns/window, whereas TI-dAFED runs give converged $\Delta F_{\rm prot}$ estimate in 5~ns/window itself.
It is noted in passing that, TI-dAFED has no additional computational cost compared to a conventional TI simulation.
%

%
%

%
%
%
%

%

\begin{figure}[htpb]
\centering
\includegraphics[width=0.8\textwidth]{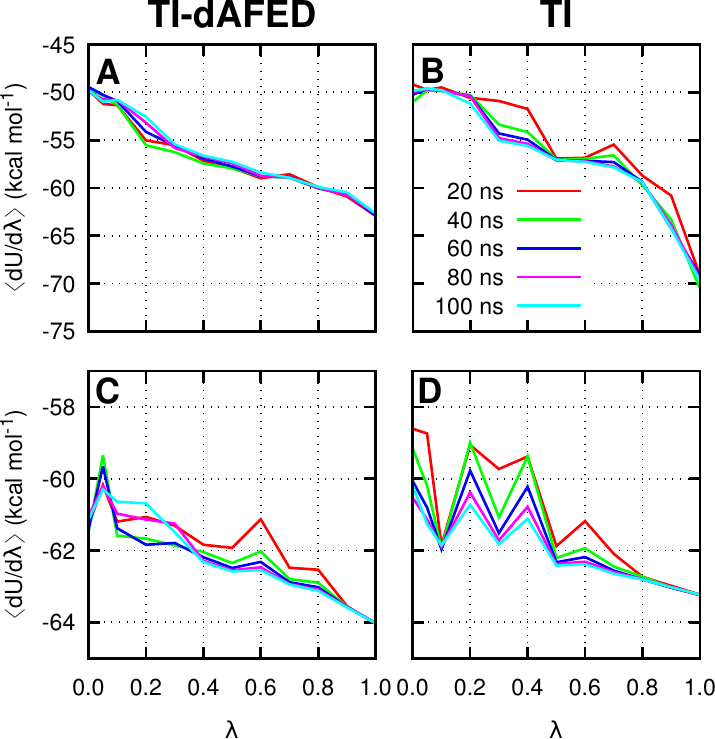}
    \caption{\label{fig:dhdl_conv} 
    Convergence of $\left < \partial U/ \partial \lambda\right >$ as a function of $\lambda$ for (A) protein using TI-dAFED method, (B) protein using TI method, (C) model using TI-dAFED method and, (D) model using TI method.
   }
\end{figure}

The convergence is examined in a more detailed manner by calculating the convergence of the derivative of free energy with respect to $\lambda$. 
\fref{fig:dhdl_conv} shows that the derivative of free energy is also well converged using both methods after 100~ns/window. %
However,  it is clear that TI-dAFED converges faster than TI for protein.
%

\begin{figure}[htpb]
\centering
\includegraphics[width=0.8\textwidth]{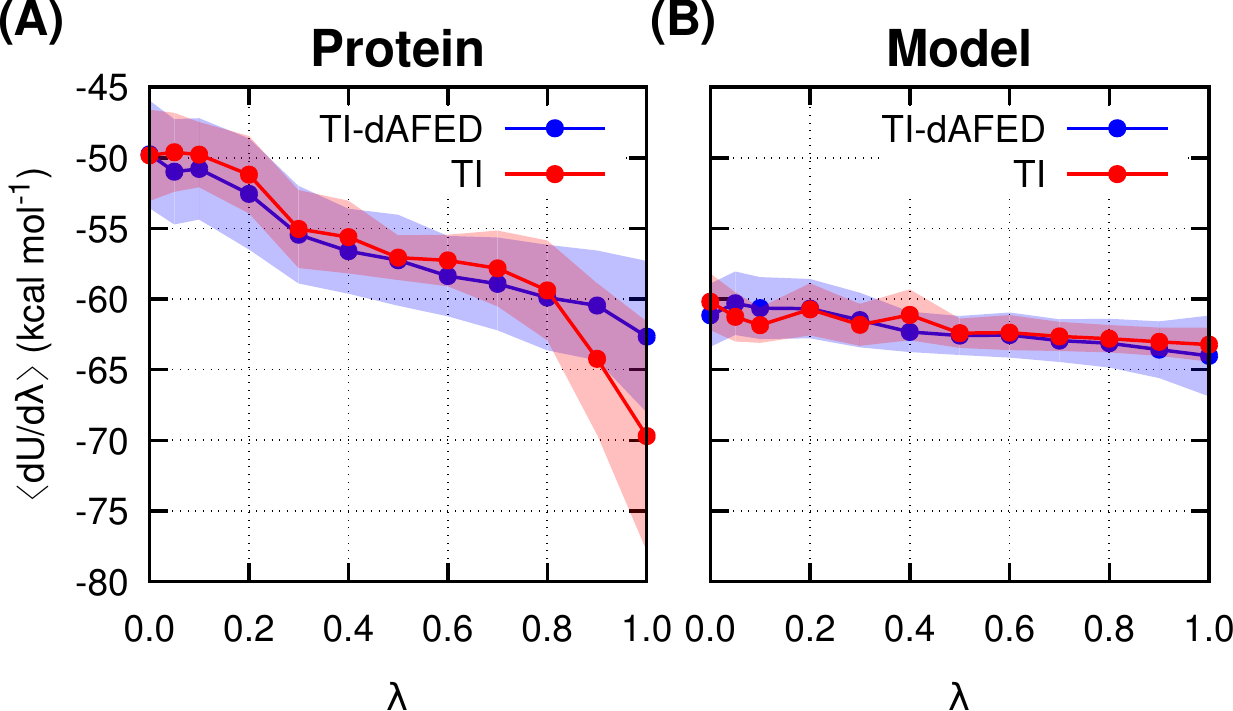}
    \caption{\label{fig:dhdl} 
    $\left < \partial U/ \partial \lambda\right >$ as a function of $\lambda$ for (A) protein and (B) model. The TI-dAFED results are in blue  and while the TI results are in red. Error bars are shown as transparent  thick lines.
    }
\end{figure}

Since we are using linear functions of $\lambda$ for $g(\lambda)$ and $f(\lambda)$, and that the electrostatic potential energy terms of Asp$_{26}$
is only varied with $\lambda$, 
$\left < \partial U/ \partial \lambda\right >$ has contributions only due to the electrostatic potential arising from the Asp$_{26}$.
%
Thus $\left < \partial U/ \partial \lambda\right >$ is ideally expected to decrease linearly with the increase in $\lambda$ from 0 to 1.\cite{Simonson2002,Simonson:JACS:2004}
%
%
%
%
Interestingly, a linear behavior of $\left < dU/d\lambda \right >$ was not found in the case of TI  for both protein and model systems, while they are nearly linear in the case of TI-dAFED simulations (\fref{fig:dhdl}).
%
%
%
%
%
%
%

%
%

\begin{figure}[htpb]
\centering
\includegraphics[width=0.8\textwidth]{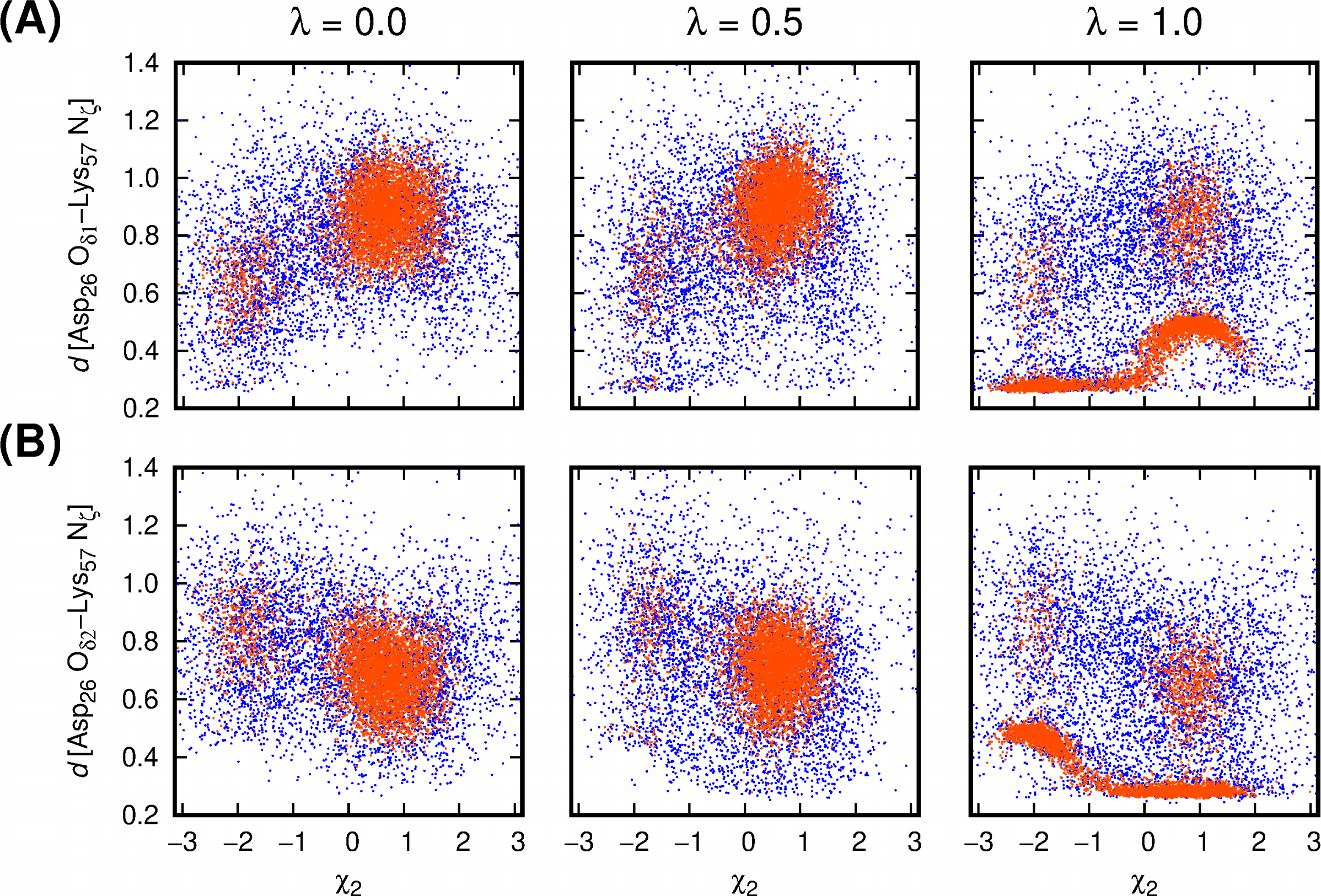}
    \caption{\label{fig:scatter_implicit} 
   Scatter plot along $\chi_2$ and $\emph{d}\, [\mathrm{Asp}_{26}\, \mathrm{O}_{\delta 1}-\mathrm{Lys}_{57} \, \mathrm{N}_\zeta$] (A) and $\chi_2$ and $\emph{d}\, [\mathrm{Asp}_{26}\, \mathrm{O}_{\delta 2}-\mathrm{Lys}_{57} \, \mathrm{N}_\zeta$] (B) with implicit solvent for $\lambda$ equals 0.0, 0.5, and 1.0. The red and the blue colors show the TI and the TI-dAFED results, respectively.
   }
\end{figure}
\begin{figure}[htpb]
\centering
\includegraphics[width=15cm,height=8cm]{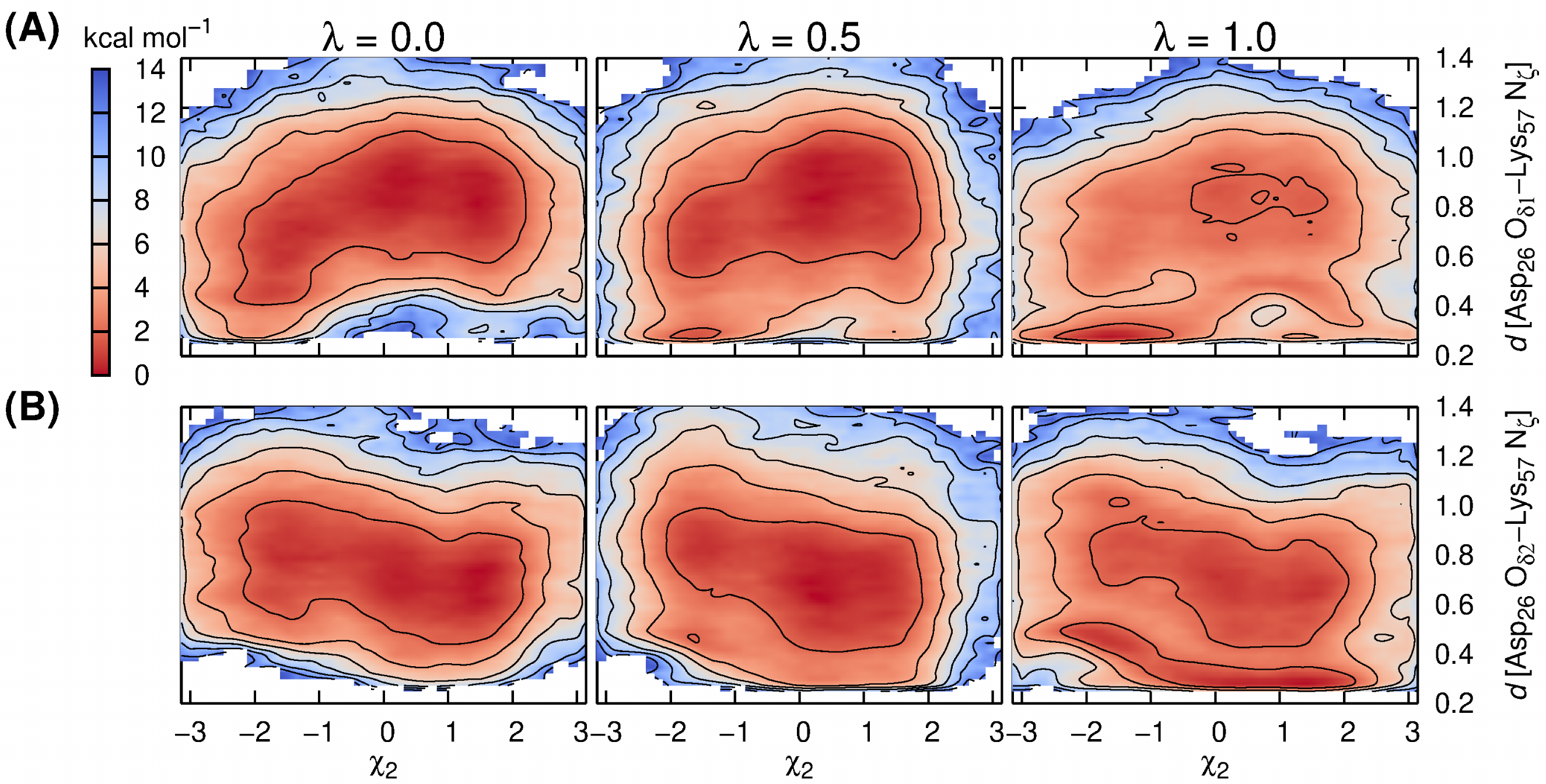}
    \caption{\label{fig:fes_implicit} 
   Free energy surface computed along $\chi_2$ and $\emph{d}\, [\mathrm{Asp}_{26}\, \mathrm{O}_{\delta 1}-\mathrm{Lys}_{57} \, \mathrm{N}_\zeta$] (A) and $\chi_2$ and  $\emph{d}\, [\mathrm{Asp}_{26}\, \mathrm{O}_{\delta 2}-\mathrm{Lys}_{57} \, \mathrm{N}_\zeta$] (B) from TI-dAFED simulations with implicit solvent for $\lambda$ values of 0.0, 0.5, and 1.0. Contours are drawn at 2~kcal~mol$^{-1}$.
   }
\end{figure}
To understand these differences, we compare the conformational sampling achieved in TI and TI-dAFED simulations.
Scatter plots of the CV values in \fref{fig:scatter_implicit} are illustrative in this respect.
Projected energy surface along the CVs for a few values of $\lambda$ are also presented in \fref{fig:fes_implicit}.
Clearly, stable basins on the free energy surfaces are visited in TI and TI-dAFED simulations.
However, within the simulation time of 100~ns, TI-dAFED simulations sample a much broader CV space than TI for
all the $\lambda$ values.
\subsection{Explicit Solvent Simulations}
%
%
The $\Delta \Delta F$ values were also computed for explicit water using TIP3P water model.
The results for the free energy differences are summarized in \tref{t:deltaF}.
The $\Delta \Delta F$ for TI-dAFED agrees with TI results.
%
However, it is 2.8 kcal mol$^{-1}$ higher than that computed using the implicit solvent model and 3.5 kcal mol$^{-1}$ higher than the experimental result.
Of great interest, an earlier simulation using explicit solvent by Simonson {\em et.~al.}~\cite{Simonson:JACS:2004} also reported a higher $\Delta \Delta F$ compared to the experimental value.

\begin{figure}[htpb]
\centering
\includegraphics[width=0.8\textwidth]{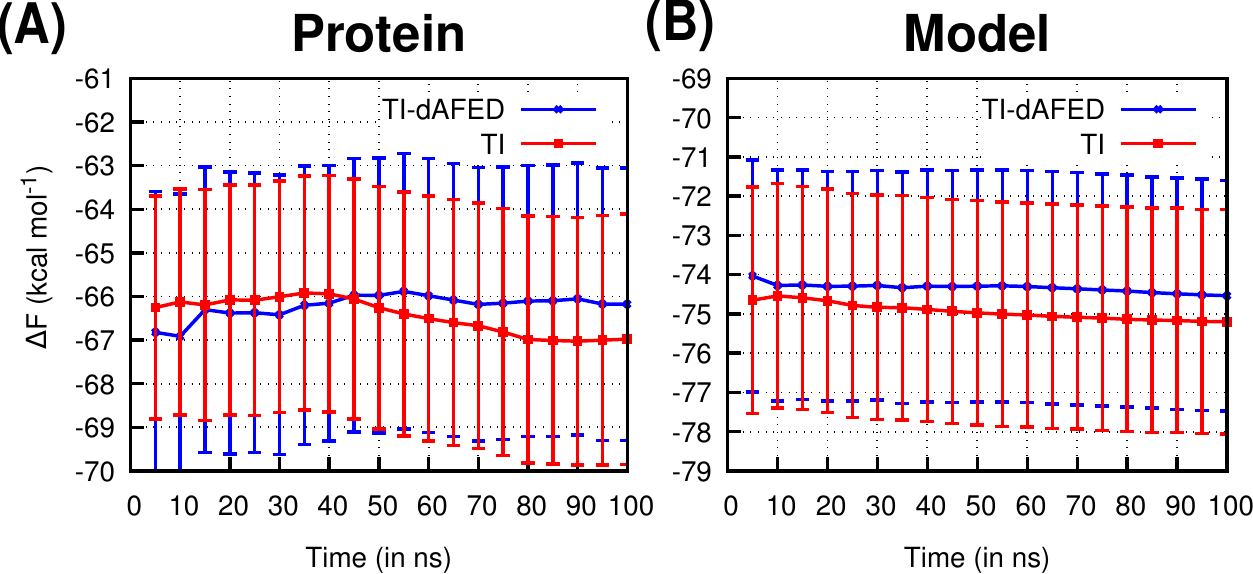}
    \caption{\label{fig:explicit_conv} 
    Convergence of $\Delta F$ in simulations using the TIP3P water model for the (A) protein and the (B) model system. Results of TI-dAFED (blue) and TI (red) are presented. Error bars are also shown.
    }
\end{figure}

\begin{figure}[htpb]
\centering
\includegraphics[width=0.8\textwidth]{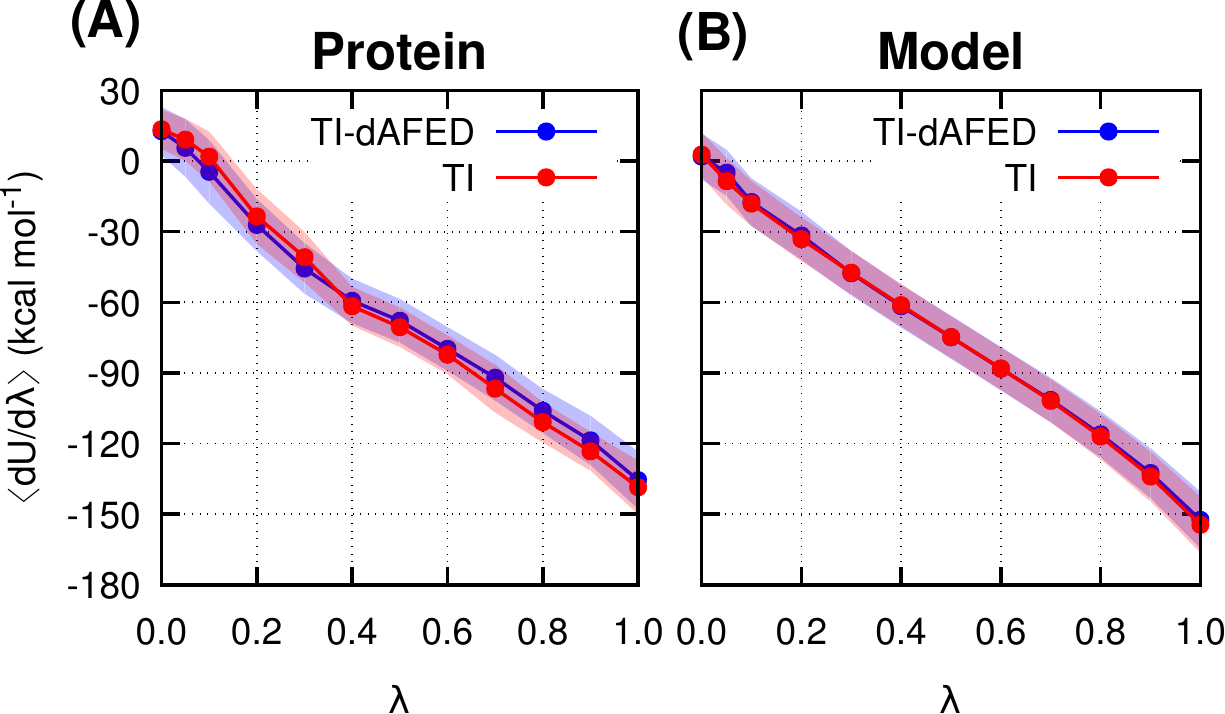}
    \caption{\label{fig:dhdl_ex} 
    $\left < \partial U/ \partial \lambda\right >$ as a function of $\lambda$ for (A) protein and (B) model. The TI-dAFED results are in blue  and while the TI results are in red. Error bars are shown as transparent  thick lines.
     }
\end{figure}

\begin{figure}[htpb]
\centering
\includegraphics[width=0.8\textwidth]{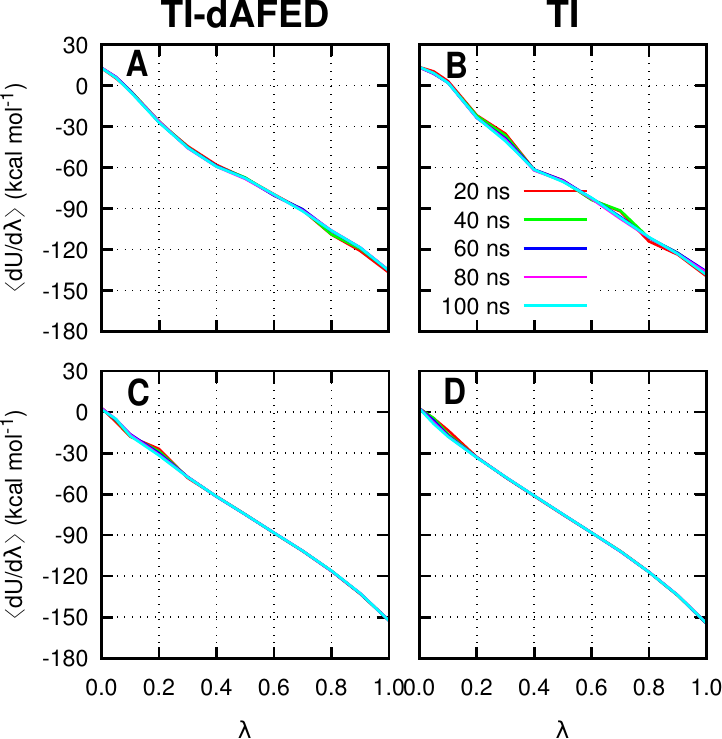}
    \caption{\label{fig:dhdl_conv_ex} 
    Convergence of $\left < \partial U/ \partial \lambda\right >$ as a function of $\lambda$ for (A) protein using TI-dAFED method, (B) protein using TI method, (C) model using TI-dAFED method and, (D) model using TI method. In these simulations, the TIP3P water model was used.
    }
\end{figure}

%
%
The convergence of $\Delta F$ and $\left < \partial U/ \partial \lambda\right >$ is shown in \fref{fig:explicit_conv} and \fref{fig:dhdl_conv_ex}.
Both the quantities are well converged within the error bars in both TI and TI-dAFED simulations. 
The $\left < \partial U/ \partial \lambda\right >$ values for TI and TI-dAFED are also comparable with each other and show a linear trend with the change of $\lambda$ (see \fref{fig:dhdl_ex}).
%
%

\begin{figure}[htpb]
\centering
\includegraphics[width=0.8\textwidth]{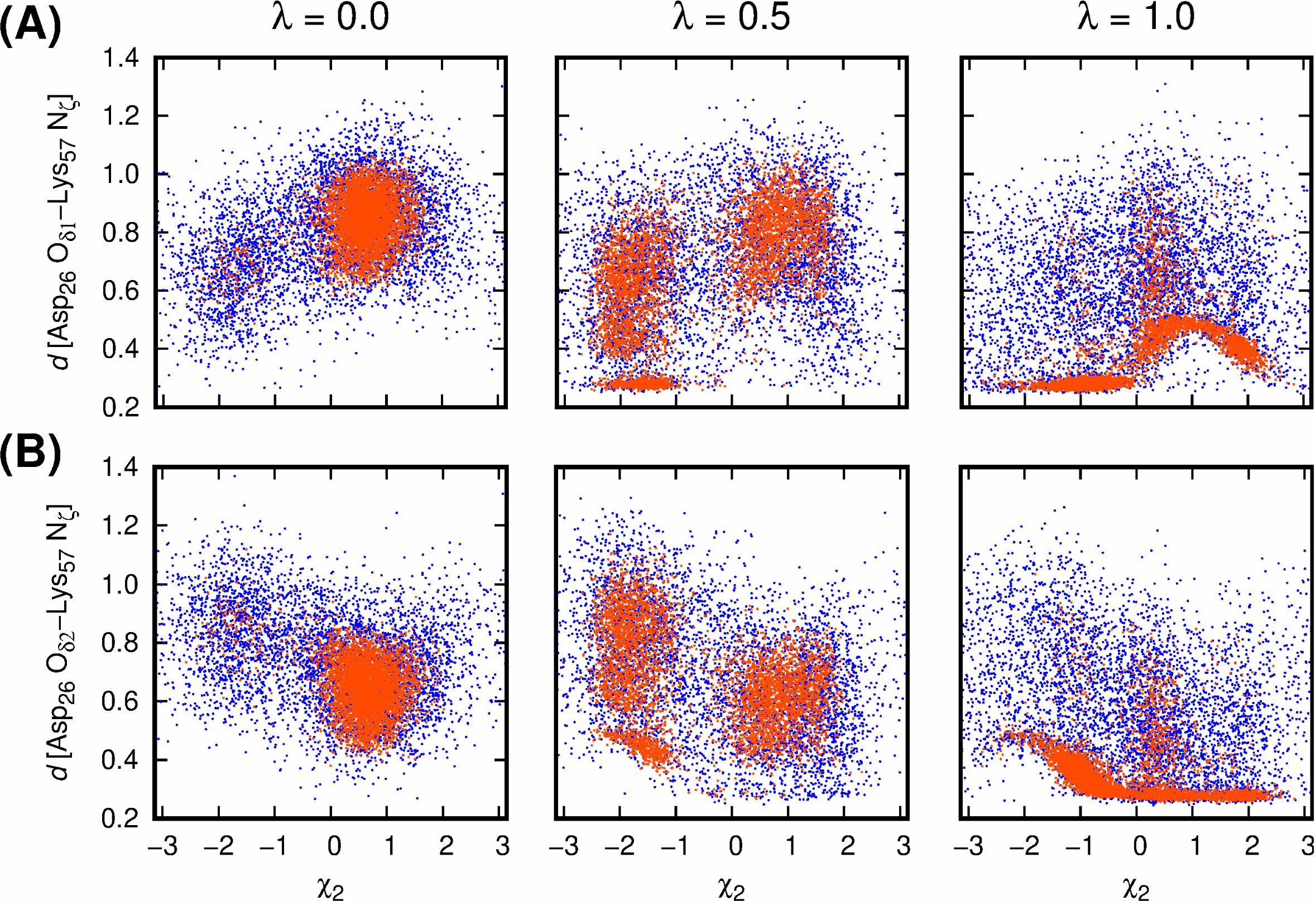}
    \caption{\label{fig:scatter_ex} 
   Scatter plot along $\chi_2$ and $\emph{d}\, [\mathrm{Asp}_{26}\, \mathrm{O}_{\delta 1}-\mathrm{Lys}_{57} \, \mathrm{N}_\zeta$] (A) and $\chi_2$ and $\emph{d}\, [\mathrm{Asp}_{26}\, \mathrm{O}_{\delta 2}-\mathrm{Lys}_{57} \, \mathrm{N}_\zeta$] coordinates (B) with TIP3P water model for $\lambda$ equals 0.0, 0.5, and 1.0. The red and the blue colors show the TI and the TI-dAFED results, respectively.
   }
\end{figure}

\begin{figure}[htpb]
\centering
\includegraphics[width=15cm,height=8cm]{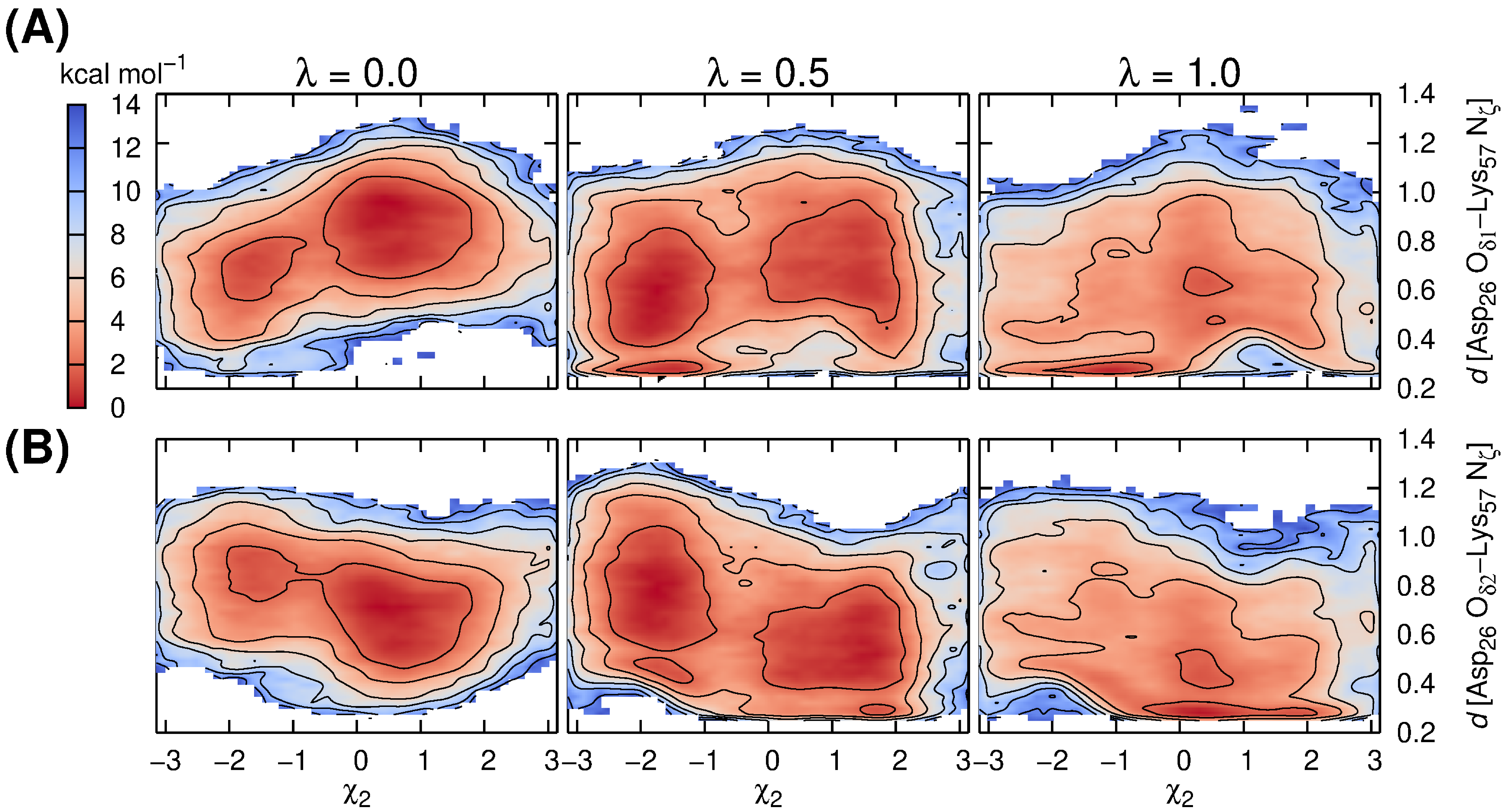}
    \caption{\label{fig:fes_ex_tip3p} 
   Free energy surface computed along $\chi_2$ and $\emph{d}\, [\mathrm{Asp}_{26}\, \mathrm{O}_{\delta 1}-\mathrm{Lys}_{57} \, \mathrm{N}_\zeta$] (A) and $\chi_2$ and  $\emph{d}\, [\mathrm{Asp}_{26}\, \mathrm{O}_{\delta 2}-\mathrm{Lys}_{57} \, \mathrm{N}_\zeta$] (B) from TI-dAFED simulations with TIP3P water model, for $\lambda$ equals 0.0, 0.5, and 1.0. Contours are drawn at 2 kcal mol$^{-1}$.
   }
\end{figure}

Like in the case of implicit solvent, we find that the conformational sampling in TI-dAFED simulation is significantly higher than TI (\fref{fig:scatter_ex} and \ref{fig:fes_ex_tip3p}), although all the minima are still sampled in TI.
%
%
%
%
%
%
%

To probe the reason for higher $\Delta \Delta F$ while using TIP3P solvent, 
we have repeated these calculations using the TIP4P water model.
%
%
The results for the free energy differences are summarized in \tref{t:deltaF}.
We found that $\Delta \Delta F$ for TI-dAFED is only 0.5~kcal~mol$^{-1}$ lesser than TI results; see also SI~Figures 1-5.

Thus we conclude that the water model is affecting $\Delta \Delta F$ estimate. 
This could be because non-polarizable TIP3P and TIP4P models may not be able to mimic the correct behavior of water molecules in the hydrophobic pocket in the vicinity of Asp$_{26}$.
%
As pointed out in the earlier works\cite{Ji2008,Burger2013}
a polarized force field might be necessary.
\section{Conclusion}
TI calculations were performed to compute $\Delta$\pka of Aps$_{26}$ in thioredoxin protein.
We reported the performance of TI-dAFED method for computing $\Delta$\pka.
It has been found that TI-dAFED can sample the conformational space exhaustively compared to conventional TI simulations.
This aids in quick convergence of $\left < \partial U/ \partial \lambda\right >$ and $\Delta F$.

The predicted value of $\Delta \Delta F$ of Aps$_{26}$ in thioredoxin protein
is in excellent agreement with the experimental data when a continuum solvent is used.
Contrarily, TIP3P and TIP4P water models are unable to provide
 a good quantitative prediction of $\Delta \Delta F$, although the direction of the shift is 
 correctly reproduced.
The differences in the $\Delta \Delta F$ between TIP3P explicit solvent simulations and the experimental data are within the error. 
%
We have found that the quantitative difference in the results is not due to the poor sampling of conformational space when an explicit solvent is taken.
Our results point out that a polarized water model may be required to capture the response 
of the changing electrostatic field around Asp$_{26}$ along with the change in $\lambda$, 
in agreement with the earlier findings.\cite{Ji2008,Burger2013}
\begin{acknowledgement}
The authors thank Prof. M. E. Tuckerman (New York University), Adrian E. Roitberg
(The University of Florida), Dr. Michel A. Cuendet (Lausanne University Hospital), and Dr. Suman Chakrabarty (S. N. Bose National Centre for Basic Sciences) for fruitful discussions.
The support of the Science and Engineering Research Board (India) under the Core Research Grant (Project No: CRG/2019/001276)
is gratefully acknowledged.  
A part of the computational resources was provided by the PARAM Sanganak supercomputing facility under the National Supercomputing Mission at IIT Kanpur.
SV thank INSPIRE (Department of Science and Technology) and IIT Kanpur for her Ph.D. fellowship.
\end{acknowledgement}

\begin{suppinfo}
The Supporting Information is available free of charge at
https://pubs.acs.org/doi/xxx. Various plots from the protein and the model-ligand simulations using TIP4P water model 
are shown: (i) convergence of  $\Delta F$,  (ii) $\left < \partial U/\partial \lambda \right >$ as a function of $\lambda$, (iii) convergence of  $\left < \partial U/\partial \lambda \right >$  as a function of simulation length, (iv) scatter plot of CVs for different values of $\lambda$, and (v) free energy surfaces along the CV space for different $\lambda$ values.
\end{suppinfo}

\bibliography{ti-tass}

\end{document}